\documentclass[12pt,preprint]{aastex}

\begin{document}

\title{The Carina-Near Moving Group}

\author{B. Zuckerman}
\affil{Dept. of Physics \& Astronomy and NASA Astrobiology Institute\\
University of California, Los Angeles\\
Los Angeles, CA 90095--1562, USA, ben@astro.ucla.edu}

\author{M. S. Bessell}
\affil{Research School of Astronomy and Astrophysics\\ 
Institute of Advanced Studies\\
The Australian National University, ACT 2611, Australia, 
bessell@mso.anu.edu.au}

\author{Inseok Song, S. Kim}
\affil{ Gemini Observatory\\
670 North A`ohoku Place\\
Hilo, HI 96720, USA, song@gemini.edu, skim@gemini.edu}

\begin{abstract}
We identify a group of $\sim$20 co-moving, mostly southern
hemisphere, $\sim$200\,Myr old stars near Earth.  Of the stars
likely to be members of this Carina-Near Moving Group, in either its
nucleus ($\sim$30\,pc from Earth) or surrounding stream, all but 3
are plausible members of a multiple star system.  The nucleus is
(coincidentally) located quite close to the nucleus of the AB
Doradus moving group notwithstanding that the two groups have
substantially different ages and Galactic space motions, UVW.
\end{abstract}


\section{Introduction}

During the past decade various young stellar associations and
clusters have been identified within 100\,pc of the Sun
\citep[see][for a review]{ZS2004}.  These comoving groups can
illuminate our understanding of the properties of low and
intermediate mass stars at times that correspond to the long-ago
era when planet building and heavy bombardment were occurring in
our Solar System.  The Spitzer Infrared Observatory is targeting
virtually all these young stars to determine if they are
surrounded by dusty debris disks.  And the stars are being
observed at near-infrared wavelengths with the Hubble Space
Telescope and with adaptive optics systems on most of the world's
largest telescopes; the goal is to detect thermal emission from
young massive planets.

In the present Letter we add another example -- the Carina-Near
Moving Group -- to the growing menagerie of nearby young
associations.  The nucleus -- consisting of stars in the relatively
narrow range of spectral type F through K0 of which HR\,3070 (=
HIP~38160) is the earliest spectral type and the brightest -- is
only $\sim$30\,pc from Earth (Table~1).  Some of the surrounding
stream stars, whose group membership is therefore less secure than
those in the nucleus, are even closer to Earth; for example,
GJ\,358 is within 10\,pc.  Based primarily on lithium abundance
and X-ray flux (see Section~\ref{sec3}), we estimate the group to be
somewhat older than the Pleiades.

Some nuclear stars in our proposed group were initially suggested
as co-moving by \citet{Eggen}. Later, \citet[hereafter
MU2000]{MU2000} argued that they are part of a spatially very
extensive, X-ray bright, moving group they dubbed ``Carina-Vela''.
However, other than X-ray luminosity, their choice of group
membership was predicated entirely on kinematics, while lacking
even radial velocity determinations.  Our past experience
\citep[][]{SBB2002} with stars proposed \citep{MF01} as members of
the TW Hydrae Association, demonstrated that a purely
convergent-point kinematic approach is insufficient to reliably
pinpoint members of nearby young associations.  Nonetheless, in
the 5 years following publication of their paper, neither
\citeauthor{MU2000} nor any other astronomer has verified or
denied their kinematic model.  For example, both \citet{C2000} and
\citet{W2003} studied group member HIP\,37563 without referencing
the MU2000 paper.  Indeed, unware of the MU2000 paper, we
independently stumbled on the nearer (to Earth) members of their
proposed association while pursuing our research program of
identification of young nearby stars. Apparently, in their young
star program, \citet{Jensen} also find some overlap with the
MU2000 sample. However, based on the few stars explicitly
mentioned in the Jensen~et~al abstract, the stars in their program
appear, on average, to be about 3 times more distant from Earth
than stars in our Table~1. 

Our investigations of radial velocities and youth indicators such
as lithium abundance and H$\alpha$ intensity are described in the
following sections.  The essence of our results may be summarized
as follows:  we confirm some of the MU2000 suggested members, deny
others, and add some of our own.  In addition, we think it is not
helpful, and may even be wrong, to associate very nearby stars
such as those listed in Table~1 with the much more distant stars
that MU2000 place together into a single moving group.
Specifically, as is evident from their Fig. 5, the nearest members
of their ``Carina-Vela'' group are three times closer to Earth
than their proposed more distant members, with essentially no
stars at intermediate distances.  In view of these large spatial
separations, lack of evidence that Galactic space velocities $UVW$
are all similar, and lack of data on lithium abundances and
activity indicators, at present we see no compelling reason to
suppose the Carina-Vela group stars shared a common origin.

Table~1 of MU2000, ``Candidate members of the Carina-Vela moving
group'', contains 58 entries most of which are not particularly
near Earth.  Of these 58, only 5 are included in our Table~1 as
candidate members of the moving group of stars we are considering
and one MU2000 candidate is a definite non-member of our proposed
moving group.  Rather than being spread out to distances
$>100$\,pc and with generally unknown $UVW$ (i.e.  MU2000's
Carina-Vela Group), our Table~1A and B stars are characterized by
proximity to Earth and rather tight constraints (a few km/s) on
measured $UVW$ velocities.  In addition, stars listed in our
Table~1B are not in the Carina-Vela direction. Finally, based on
our Figures 1 and 2 and the discussion in Section 3, we believe
that the Carina-Near stars must be older than the Pleiades,
whereas IC\,2391 which belongs to the Carina-Vela moving group
proposed by MU2000 is generally regarded as younger than the
Pleiades. In view of these many differences, we designate the
stars we are considering to be the ``Carina-Near Moving Group''.

\section{Observations}

In January 2006, we obtained spectra of Table~1 stars with the
double-beam grating and echelle spectrographs on the two Nasmyth
foci of the Australian National University's 2.3\,m telescope at
Siding Spring Observatory.  The primary goal of these observations
is the measurement of stellar radial velocity along with the
equivalent widths of the H$\alpha$ and Li\,$\lambda$ 6708 lines.
Radial velocity, in conjunction with proper motion and parallax,
enables one to calculate the three-dimensional Galactic space
motions U,V,W that are essential for identification of individual
moving groups.

Entries for radial velocity, lithium, and H$\alpha$ in
Table~1 are a mix of our own measurements and those of others.
U,V,W was calculated using weighted averages of proper motions
from the PPM and Tycho catalogs and occasionally a third catalog
and of radial velocities measured by us and by others. For stars
that are obviously binaries based on their small separation in the
plane of the sky, for example HIP\,37918 and 37923, we used the
weighted mean of their Hipparcos measured distances to calculate
the UVWs listed in Table~1.

We have divided Table~1 into three parts: (A) nuclear members of
the Carina-Near Moving Group, (B) suggested ``stream'' members
associated with the Group, and (C) stars suggested in MU2000 as
comoving but rejected by us as group members based on UVW.
For moving groups with ages
$<50$\,Myr, mid and late M-type stars located above the main
sequence are probably the most reliable way to establish age 
\citep[see, e.g., discussion and Fig. 2 in][]{ZS2004}.  However,
we could identify no stars noticeably above the main sequence
among our proposed group members.  The absence of such pre-main
sequence stars is consistent with our $\sim200$\,Myr age estimate
for the group (Section~\ref{sec3}).

\section{Discussion\label{sec3}}

\begin{figure}
\begin{center}
 \includegraphics[width=0.95\columnwidth]{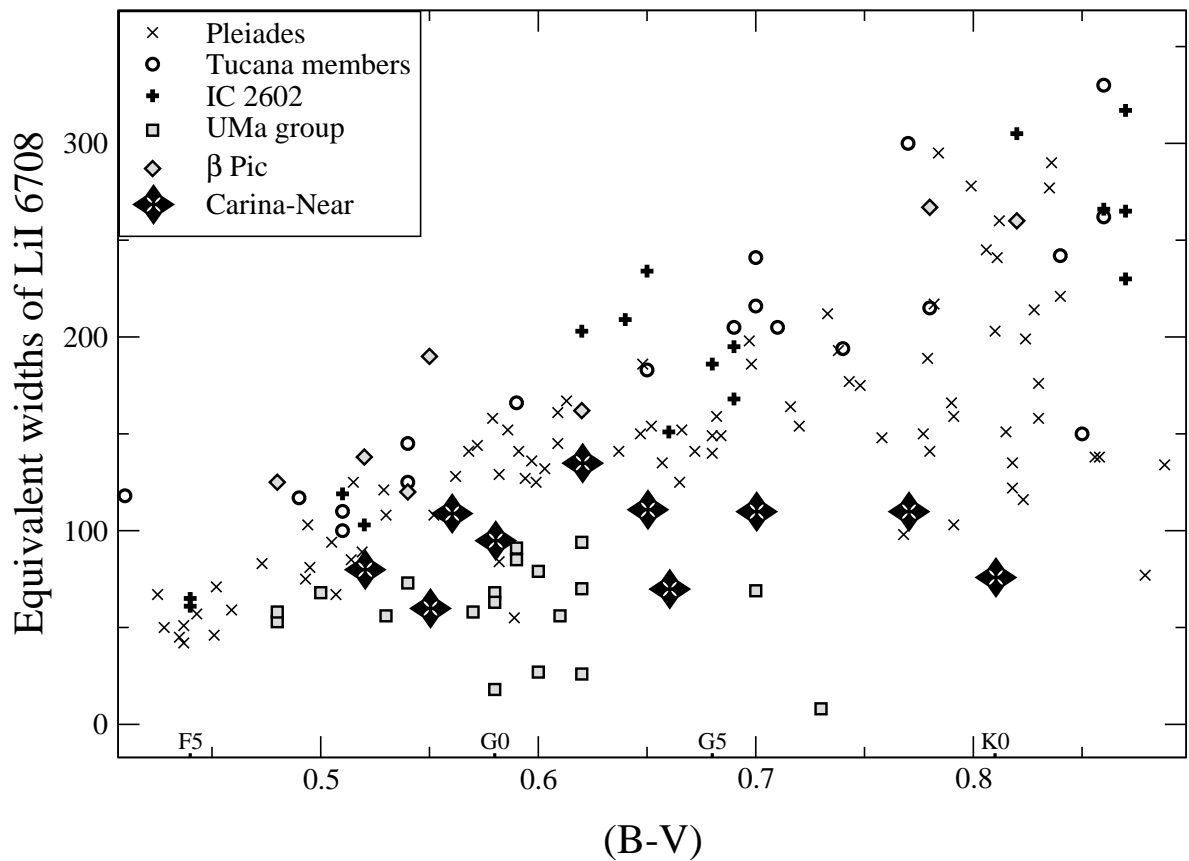}
\end{center}
\caption{Equivalent width of Li\,I\,6708\,\AA\ as a function of
$B-V$. Displayed equivalent widths are not corrected for possible
contamination by Fe\,I\,6707.44\,\AA, and measurement uncertainty
of equivalent widths is $\sim20$\,m\AA.  $B-V$ values for Carina-Near
group stars are calculated from Tycho-2 $B_T$ and $V_T$ magnitudes
using the relation given in \citet{Bessell}.\label{Lithium}}
\end{figure}

\begin{figure}
\begin{center}
 \includegraphics[width=0.95\columnwidth]{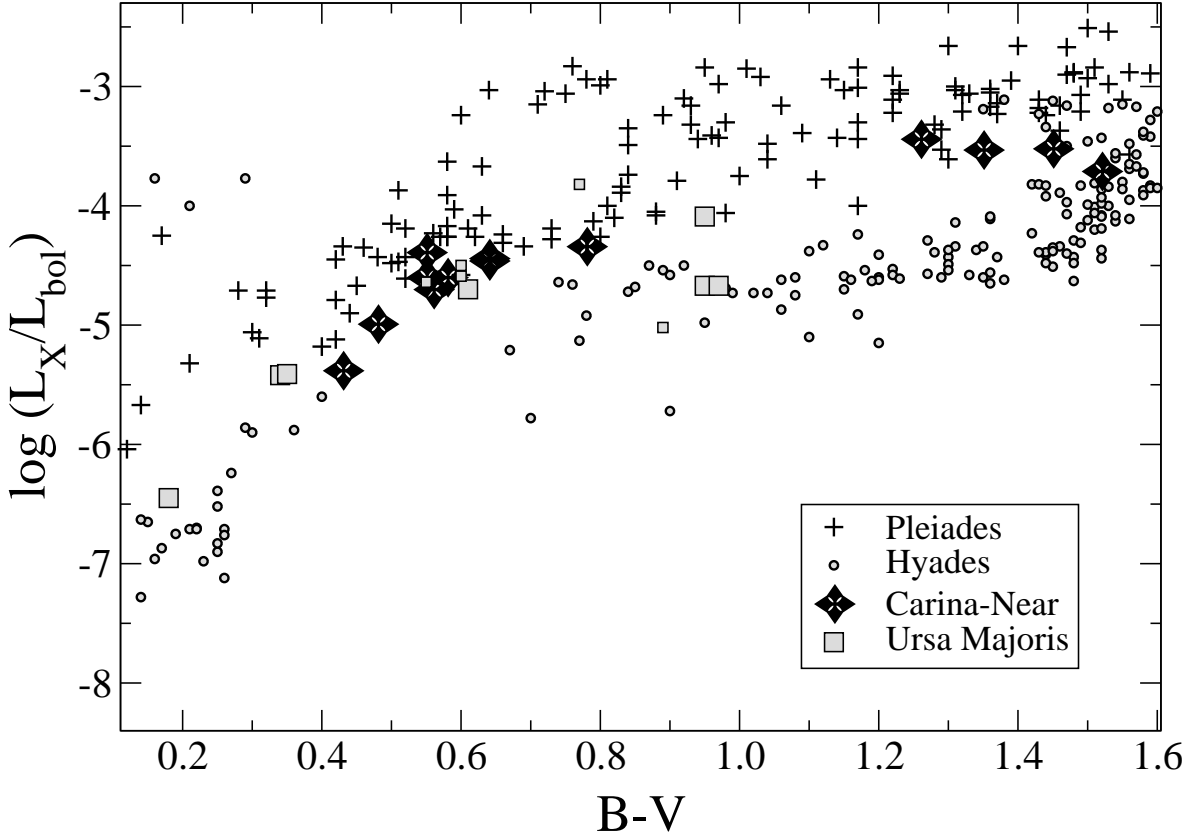}
\end{center}
\caption{Ratio of X-ray to bolometric luminosity as a function
$B-V$. $B-V$ values for Carina-Near group stars are calculated from
Tycho-2 $B_T$ and $V_T$ magnitudes as in Figure~1. Ursa Majoris
moving group (estimated age of $500\pm100$\,Myr; \citealt{King})
stars are plotted as square symbols (large squares for nucleus
members). Based on the relative position in this plot of UMa and
Hyades member stars, if the Hyades age is $\sim600$\,Myr as is
usually quoted, then the UMa stars are probably $\sim400$\,Myr old.
B-V color of a multiple system was calculated to reproduce single 
 color for matching the composite  Lx/Lbol value of the system.
\label{Xray}}
\end{figure}

Lacking any M-type stars that lie above the main sequence, we
estimate the age of the Carina-Near group using lithium abundance
plus activity indicators.  Figure~\ref{Lithium} displays
Li\,$\lambda$6708 equivalent width (EW) vs.  spectral type of
Carina-Near group stars overplotted on Fig. 3 from \citet{ZS2004}.
The Carina-Near stars are seen to lie near the bottom envelope of
lithium EW for Pleiades stars.  Figure~\ref{Xray} shows the X-ray
luminosity of Carina-Near stars, normalized by their bolometric
luminosities, overplotted on Fig. 4 of \citet{ZS2004}.  Again, the
Carina-Near stars lie near the bottom envelope of Pleiades stars.
Finally, we measured H$\alpha$ lines in early-M type stars
HIP\,47425 and   15844AB and C (all of which have $B-V\sim1.5$),
to be $-0.77$,$-0.14$ and $+0.29$   respectively where a negative
sign denotes emission.  Comparison with H$\alpha$ line activity in
stars with similar $B-V$, see Fig.~5 in \citet{ZS2004} and Fig. 5
in \citet{G2002}, suggests that these stars are young, but not
dramatically so.

With adaptive optics imaging, \citet{M2003} found that GJ\,900
(HIP\,116384), is a triple system.  Martin notes that
his referee, J.  Stauffer, argues that the age of GJ900 is
100\,Myr or older based on its  position in a $V$ vs $V-I$
color-magnitude diagram and comparison with   low-mass members of
the IC\,2391, IC\,2602, and Pleiades clusters.

Our measurements, as indicated in Figures 1 and 2, are consistent
with Stauffer's assessment and we assign an age of $200\pm50$\,Myr
to the Carina-Near Moving Group.

A remarkable aspect of the stars listed in parts (A) and (B) of
Table 1 is the dominance of multiple star systems.  Only
HIP\,36414, 37635 and 47425 appear, at this time, to be likely
single stars.  The situation regarding HIP\,37563 and 38160 is
unclear.  While HIP\,37563 and 38160 might be single stars (they
are fairly far from each other in the plane of the sky), it is
possible that either one or both might comprise a very wide
multiple system with binary star HIP\,37918 \& 37923.
Specifically, HIP\,37563 and 38160 are each separated in the plane
of the sky by about one light year from HIP\,37918 \& 37923.  If
the separation along the line of sight is comparable (or less),
then HIP\,37563 or 38160 or both could, with 37918 \& 37923, be a
very wide triple or quadruple system.

In their Section 5, MU2000 comment that ``surprisingly many'' of
their proposed Carina-Vela group stars are members of binaries;
this trait, they note, may be characteristic of stellar youth. If
so, then such youthful affinity for multiplicity has been
preserved for $\sim200$\,Myr in the Carina-Near group.

\section{Conclusions}

We have identified a group of comoving $\sim$200\,Myr old stars
that partially surround the Sun.  At least four of these stars are
close enough to Earth to have been included in the original Gliese
catalog of nearby stars (1969):  GJ\,140, 358, 900, and 907.1.
Like all previously discovered and younger associations of stars
within 100\,pc of Earth -- TW~Hydrae, Tucana/Horologium,
$\beta$~Pictoris, AB~Doradus, $\eta$~Cha, Cha-Near -- the nucleus
of the Carina-Near Moving Group is located deep in the southern
hemisphere.  This nucleus, is (coincidentally) located quite close
to the nucleus of the AB~Doradus Moving Group notwithstanding that
the two groups have substantially different ages and Galactic
space motions, UVW: AB~Dor $(-8,-27,-14)$; Carina-Near
$(-26,-18,-2)$. Indeed, the U velocity of the Carina-Near group is
substantially more negative than the U velocities of all the other
groups listed here.  In any event, identification of this nearby
group of stars, many of which have masses similar to the Sun,
extends the time-line covered by these other groups to older ages.
As instrumentation capabilities improve, astronomers will be able
to follow the planetary formation process over the age interval
from 8 to 200\,Myr.

\acknowledgements{We thank an anonymous referee and referee
M.Schmitz for constructive comments.}

\clearpage

\begin{deluxetable}{cccccrrrr@{$\pm$}llrrc}
\def\mc{\multicolumn}
\def\ch{\colhead}
\def\mnodata{\mc{2}{c}{\nodata}}
\setlength{\tabcolsep}{0.02in}
\tablecolumns{13}
\tablewidth{0pc}
\tabletypesize{\scriptsize}
\tablecaption{The Carina-Near Moving Group}
\tablehead{
\ch{HIP}&\ch{Other}&\ch{R.A.}   &\ch{Dec.}   &\ch{Sp.}  & \ch{V}&\ch{B-V} & \ch{D}   &\mc{2}{c}{RV}    &\ch{Lx/Lbol}&\ch{Li}  &\ch{H$\alpha$}&\ch{U,V,W (uncertainty)}  \\
        &\ch{Name} &\ch{(J2000)}&\ch{(J2000)}&\ch{Type} &\ch{(mag)}&     &\ch{(pc)}&\mc{2}{c}{(km/s)}&\ch{(log)}  &\ch{(m\AA)}&\ch{(\AA)}   &\ch{(km/s)}  }
\startdata
\cutinhead{A. Members of the nucleus of the Carina-Near Moving Group}
 35564A & \nodata &07 20 21.4&-52 18 41 &  F2  &  6.05 & 0.48 & 34.8 &18.6 &  2  & -4.99$^\dagger$ &\nodata&   1.8 &  $-26.2 -15.3  -4.6$(  1.2  1.9  1.1) \\
 35564B & \nodata &07 20 21.9&-52 18 33 &  G0  &  6.60 & 0.58 & 34.8 &\mnodata   & -4.99$^\dagger$   &   95  &   1.0 &  \nodata                              \\
 36414  & HD59704 &07 29 31.4&-38 07 21 &  F7  &  7.75 & 0.52 & 52.5 &  28 &  2  & -4.39   &   80  &   1.3 &  $-26.1 -20.7  -2.6$(  0.9  1.9  0.5) \\
 37563  & HD62850 &07 42 36.1&-59 17 51 &  G2/3&  7.19 & 0.62 & 33.3 &  17 &  1  & -4.44   &  135  &   1.1 &  $-25.5 -18.0  -1.8$(  0.5  1.0  0.4) \\
 37635  & HD62848 &07 43 21.5&-52 09 51 &  G0  &  6.70 & 0.55 & 30.2 & 20.5& 0.5 & -4.60   &   60  &   1.2 &  $-25.3 -18.4  -2.5$(  0.4  0.5  0.2) \\
 37918  & HD63581 &07 46 14.8&-59 48 51 &  K0  &  8.15 & 0.77 & 36.2 &  17 &  1  & -4.34$^\dagger$   &  110  &   0.9 &  $-26.0 -18.1  -2.4$(  1.8  1.0  0.4) \\
 37923  & HD63608 &07 46 17.0&-59 48 34 &  K0  &  8.25 & 0.81 & 27.4 &  17 &  1  & -4.34$^\dagger$   &   76  &   0.9 &  $-26.0 -18.1  -2.4$(  1.8  1.0  0.4) \\
 38160  & HR3070  &07 49 12.9&-60 17 01 &  F1  &  5.79 & 0.43 & 34.9 &  15 &  2  & -5.38   &\nodata&   2.1 &  $-26.2 -16.6  -1.6$(  0.5  1.9  0.6) \\
\cutinhead{B. Probable and possible members of the Carina-Near Stream}
 15844AB& GJ140   &03 24 06.5&+23 47 06 &  M1  & 10.4  & 1.45 & 19.8 &  19 &  2  & -3.52   &    0  &  -0.1 &  $-24.3 -17.0  -6.0$(  1.9  2.1  0.9) \\
 15844C & GJ140C  &03 24 12.7&+23 46 20 &  M2.5& 11.9  &\nodata&19.8 & 18  & 3   & \nodata &    0  &  +0.3 &  $-23.3 -15.0 - 4.6$(  2.7  2.0  1.4) \\
 47425  & GJ358   &09 39 46.4&-41 04 03 &  M2  & 10.77 & 1.52 &  9.5 &  18 &  3  & -3.71   &    0  &  -0.8 &  $-28.8 -17.8  -0.9$(  0.5  3.0  0.5) \\
 58240  & HD103742&11 56 42.3&-32 16 05 &  G3  &  7.64 & 0.65 & 34.9 &   6 & 0.4 & -4.46$^\dagger$   &  111  &   1.1 &  $-21.2 -17.1  -3.1$(  3.3  1.8  0.9) \\
 58241  & HD103743&11 56 43.8&-32 16 03 &  G3  &  7.81 & 0.70 & 29.2 &  6.7& 0.3 & -4.46$^\dagger$   &  110  &   1.1 &  $-21.0 -17.6  -2.8$(  3.3  1.8  0.9) \\
 60831  & HD108574&12 28 04.4&+44 47 39 & (F7) &  7.40 & 0.56 & 39.2 & -1.8&  1  & -4.70$^\dagger$   &  109  &\nodata&  $-28.5 -17.8  -4.4$(  2.0  1.2  1.0) \\
 60832  & HD108575&12 28 04.8&+44 47 30 & (G0) &  7.97 & 0.66 & 42.4 &\mnodata   & -4.70$^\dagger$   &   70  &\nodata&  $-30.9 -18.3  -1.5$(  4.9  3.0  2.0) \\
 116384 & GJ900   &23 35 00.3&+01 36 19 &  M0  &  9.61 & 1.35 & 19.3 &  -10&  2  & -3.53   &$6\pm2$&\nodata&  $-29.1 -15.7  +0.9$(  1.0  1.2  1.7) \\
 117410 & GJ907.1 &23 48 25.7&-12 59 15 &  K8  &  9.57 & 1.26 & 27.1 &  -8 &  2  & -3.44   &\nodata&\nodata&  $-29.0 -14.0  +0.9$(  1.8  1.0  1.9) \\
\cutinhead{C. Non-members}
 37718  & HD63008 &07 44 12.5&-50 27 24 &  F8  &  6.64 & 0.53 & 30.8 &  8.5& 0.3 & -4.60$^\dagger$   &   67  &   1.3 &  $-26.4  -4.4  -6.2$(  0.3  0.3  0.2) \\
 37727  & HD63008B&07 44 16.5&-50 28 00 &  G0  &  7.55 & 0.70 & 30.1 &  9.0& 0.5 & -4.60$^\dagger$   &   46  &   1.0 &  $-26.5  -4.9  -6.4$(  0.3  0.5  0.2) \\
\hline
\enddata
\tablecomments{
$UVW$ are defined with respect to the Sun, with $U$ positive
toward the Galactic Center, $V$ positive in the direction of
Galactic rotation, and $W$ positive toward the north Galactic
pole.\\
Average UVW for the nucleus of the Carina-Near Moving Group is $-25.9,-18.1,-2.3$\,(km/sec).\\
HIP 35564B:  We measured the radial velocity to be 59\,km/sec and 32\,km/sec
in 2006 January and April, respectively; so this star may be a
spectroscopic binary.\\
HIP\,37563:  We measured 125\,m$\AA$ for lithium 6708\,\AA.  Wichmann et al (2003) 
                and Cutispoto et al (2002) each measured 145\,m\AA.\\
HIP\,37718 \& 37727:  parallax used for calculation of UVW is $32.8\pm0.4$~mas. 
             Proper motion = ($-110.8\pm1.0, +143.0\pm1.0$) mas/yr \\
HIP\,37918 \& 37923:  parallax used is $29.2\pm2.0$~mas.  
             Proper motion = ($-58.0\pm1.5, +155\pm1.5$) mas/yr\\
HIP\,47425 : \citet{G-S} list RV$=142\pm21$\,km/sec.\\
HIP\,58240 \& 58241: parallax used is $31.45\pm4.5$~mas.  
             Proper motion = ($-177.4\pm2.0, -5.3\pm2.0$) mas/yr\\
HIP\,60831 \& 60832: listed spectral types are estimated from $V-K$ color.\\
HIP\,116384:  listed lithium 6708\,\AA\ equivalent width is from \citet{ZBR}. Triple system \citep{M2003}.\\
HIP\,117410: 1$''$ Hipparcos binary. \\
$\dagger$: ROSAT X-ray positional uncertainty circle covers both stars in the 4
indicated binary systems.  To calculate fractional X-ray luminosity in each
case, we ``shared'' the X-ray flux between the two stars.
}
\end{deluxetable}


\clearpage

\clearpage

\begin{thebibliography}{11}
\expandafter\ifx\csname natexlab\endcsname\relax\def\natexlab#1{#1}\fi

\bibitem[\protect\astroncite{{Bessell}}{2000}]{Bessell}
{Bessell}, M.~S. 2000, {\em \pasp\/}, {\bf 112}, 961

\bibitem[\protect\astroncite{{Cutispoto} {\em et~al.\/}}{2002}]{C2000}
{Cutispoto}, G., {Pastori}, L., {Pasquini}, L., {de Medeiros}, J.~R.,
  {Tagliaferri}, G., \& {Andersen}, J. 2002, {\em \aap\/}, {\bf 384}, 491

\bibitem[Eggen(1988)]{Eggen} Eggen, O.~J.\ 1988, \apss, 142, 145

\bibitem[\protect\astroncite{{Garc{\'{\i}}a-S{\'a}nchez} {\em
  et~al.\/}}{2001}]{G-S}
{Garc{\'{\i}}a-S{\'a}nchez}, J., {\em et~al.\/} 2001, {\em \aap\/}, {\bf 379},
  634

\bibitem[\protect\astroncite{{Gizis} {\em et~al.\/}}{2002}]{G2002}
{Gizis}, J.~E., {Reid}, I.~N., \& {Hawley}, S.~L. 2002, {\em \aj\/}, {\bf 123},
  3356

\bibitem[Jensen et al.(2004)]{Jensen} Jensen, E.~L.~N., 
Schlesinger, K.~J., \& Higby-Naquin, C.~T.\ 2004, BAAS, 205

\bibitem[King et al.(2003)]{King} King, J.~R., Villarreal, 
A.~R., Soderblom, D.~R., Gulliver, A.~F., \& Adelman, S.~J.\ 2003, \aj, 
125, 1980

\bibitem[\protect\astroncite{{Makarov} \& {Fabricius}}{2001}]{MF01}
{Makarov}, V.~V. \& {Fabricius}, C. 2001, {\em \aap\/}, {\bf 368}, 866

\bibitem[\protect\astroncite{{Makarov} \& {Urban}}{2000}]{MU2000}
{Makarov}, V.~V. \& {Urban}, S. 2000, {\em \mnras\/}, {\bf 317}, 289

\bibitem[\protect\astroncite{{Mart{\'{\i}}n}}{2003}]{M2003}
{Mart{\'{\i}}n}, E.~L. 2003, {\em \aj\/}, {\bf 126}, 918

\bibitem[\protect\astroncite{{Song} {\em et~al.\/}}{2002}]{SBB2002}
{Song}, I., {Bessell}, M.~S., \& {Zuckerman}, B. 2002, {\em \aap\/}, {\bf 385},
  862

\bibitem[\protect\astroncite{{Wichmann} {\em et~al.\/}}{2003}]{W2003}
{Wichmann}, R., {Schmitt}, J.~H.~M.~M., \& {Hubrig}, S. 2003, {\em \aap\/},
  {\bf 399}, 983

\bibitem[\protect\astroncite{{Zboril} {\em et~al.\/}}{1997}]{ZBR}
{Zboril}, M., {Byrne}, P.~B., \& {Rolleston}, W.~R.~J.~R. 1997, {\em \mnras\/},
  {\bf 284}, 685

\bibitem[\protect\astroncite{{Zuckerman} \& {Song}}{2004}]{ZS2004}
{Zuckerman}, B. \& {Song}, I. 2004, {\em \araa\/}, {\bf 42}, 685

\end{thebibliography}
\end{document}